# Spectrally and Spatially Configurable Superlenses for Optoplasmonic Nanocircuits

Svetlana V. Boriskina and Björn M. Reinhard
 Department of Chemistry & The Photonics Center, Boston University, Boston, MA, 02215.
sboriskina@gmail.com, bmr@bu.edu

**Abstract**
Energy transfer between photons and molecules and between neighboring molecules is ubiquitous in living nature, most prominently in photosynthesis. While energy transfer is efficiently utilized by living systems, its adoption to connect individual components in man-made plasmonic nanocircuits has been challenged by low transfer efficiencies which motivate the development of entirely new concepts for energy transfer. We introduce herein optoplasmonic superlenses that combine the capability of optical microcavities to insulate molecule-photon systems from decohering environmental effects with the superior light nanoconcentration properties of nanoantennas. The proposed structures provide significant enhancement of the emitter radiative rate and efficient long-range transfer of emitted photons followed by subsequent re-focusing into nanoscale volumes accessible to near- and far-field detection. Optoplasmonic superlenses are versatile building blocks for optoplasmonic nanocircuits and can be used to construct "dark" single molecule sensors, resonant amplifiers, nanoconcentrators, frequency multiplexers, demultiplexers, energy converters and dynamical switches.

**Introduction**
Non-radiative energy transfer between nanoobjects is limited to distances of only a few nanometers, making photons the most attractive long-distance signal carriers. However, once the photon is emitted by a donor quantum emitter, the probability of acceptor absorbing its energy becomes negligibly small. Therefore, realizing efficient and controllable on-chip interactions between single photons and single quantum emitters, which are crucial for single-molecule optical sensing and quantum information technology, remains challenging. This problem is mitigated by optical microcavities (OM), which can significantly boost the probability of a photon re-absorption through acceptor molecules (1) via efficient trapping and re-circulating of photons (2). OMs also strongly modify radiative rate of emitters at select frequencies corresponding to cavity modes, which can provide local density of optical states (LDOS) exceeding that of the free space by orders of magnitude (2-5). In turn, noble-metal nanostructures can enhance emission of free-space photons by excited molecules (effectively acting as nano-analogs of radio antennas) (6-12) or facilitate relaxation by coupling to surface plasmons (SPs) (13-15). Consequently, both plasmonic nanostructures and OMs can modify the LDOS (16, 17), but the OM approach suffers from limited accessibility of the intracavity volume by target molecules (which should either be incorporated into the cavity material (3, 4) or interact with photonic modes via their weak evanescent tails (5, 18-20)), while high dissipative losses in metals create fundamental limitations for long-distance energy and information transfer through surface plasmons (21).

**Results and Discussion**
In this paper, we develop a new approach for photon generation and energy transfer in optoplasmonic circuits that combines sub-wavelength confinement of electromagnetic fields near plasmonic nanoantennas with long photon dwelling times provided by high-Q OMs and thus achieves cascaded photon-emitter interactions over long (up to hundreds of microns) length scales. To demonstrate a general physical concept rather than optimized engineering solutions, we consider model structures composed of spherical micro- and nanoparticles and analyze them within the framework of the generalized Mie theory (GMT) (detailed in Methods) (22-24). One possible realization of an optoplasmonic superlens is shown in Fig. 1a and consists of two Au nanodimer antennas (10, 11) coupled to OM via nanoscale-size gaps. The structure is excited by the electric field $\mathbf{E}(\mathbf{r})e^{i\omega t}$ of a donor dipole source with the transition moment $\mathbf{p}$, which is centered in the gap of one of the nanodimers and serves as a model of a quantum emitter (e.g. atom, molecule or quantum dot (QD)). The donor dipole can lose its energy either radiatively by emitting a free-space photon or non-radiatively through dissipation in metal, and, within the validity of the Fermi's golden rule, its total decay rate can be expressed as a weighted sum of possible decay channels. The changes in the LDOS at the dipole position induced by the optoplasmonic superlens are used to re-distribute the "weights" of available channels. The resulting modification of the dipole radiative $\gamma_r$ and non-radiative $\gamma_{nr}$ rates can be obtained via classical calculations of the electromagnetic fields (7, 15, 25) (see Methods).

The dipole radiative rate enhancement is calculated by integrating the power flux through the closed surface encompassing the emitter and the superlens $\gamma_r = 1/2 \operatorname{Re} \oint_S \left( \mathbf{E}(\mathbf{r}) \times \mathbf{H}^*(\mathbf{r}) \right) \cdot \mathbf{n} dr$ ($\mathbf{n}$ is a unit vector normal to the surface) and normalizing to the power $\gamma_r^0$ radiated by the same source in vacuum. The polarization of the donor dipole is chosen to be oriented along the dimer axis as this orientation yields the dominant contribution to the radiative rate enhancement (Fig. 1a). Fig. 1b shows that the presence of the superlens yields two-orders-of-magnitude resonant enhancement of the dipole radiative rate. The resonant peaks in Fig. 1b are a manifestation of the excitation of the high-Q whispering-gallery (WG) modes in the OM (see SI Figs. S1,2), when photons are temporarily trapped inside the microcavity by total internal reflection. The 'acceptor' nanoantenna coupled to the opposite side of the microcavity provides a well-defined output channel, which dominates all other channels of light out-coupling via evanescent leakage through the cavity walls. The localized plasmon oscillations induced in the acceptor antenna provide both dramatic field enhancement and light localization (Fig. 1c-e). The electric field intensity distributions around the superlens demonstrate that the majority of the light emitted by the dipole is captured by the superlens and re-localized in the acceptor nanoantenna (Fig. 1d,e). In the absence of the OM the electric field intensity induced on the acceptor antenna drops four orders of magnitude, ruling out the possibility that it is directly absorbing the far-field radiation (see Fig. 1f,g and SI Fig. S3). Furthermore, plasmonic nanoantennas enable the detection of Raman radiation scattered by single molecules owing to the extreme concentration of the intensity of both the incident light and the Raman radiation in the form of localized surface plasmons (26-29). Optoplasmonic superlenses can be configured to further amplify the pump and Raman intensity, to capture the Raman-scattered light in the form of OM-trapped photons, and to subsequently re-focus into another plasmonic nanoantenna.

In the configuration shown in Fig. 1a, the SP oscillations localized on the acceptor nanoantenna are converted into free photons that can be collected by conventional far-field optics. Other detection modalities are, however, available as well. For example, locally-addressable on-chip electrical detection of surface plasmons has already been successfully demonstrated by using germanium wires (30), gallium arsenide structures (31), organic photodiodes (32), and superconducting single-photon detectors (33). Finally, light trapped in the optical microcavity can be evanescently out-coupled into an optical fiber (34), a planar optical waveguide (35, 36) or into another microcavity resonator (24, 37). The combination of dramatic field nanoconcentration and cascaded signal enhancement in optoplasmonic superlenses with the possibility of the on-chip routing and detection of the amplified signal paves the way for the realization of sensitive "dark" optoplasmonic platforms for single-molecule detection.

While an individual optoplasmonic superlens can serve as a super-resolution magnifying glass for the investigation of nanoscale objects, additional functionality can be obtained by integrating the superlenses into discrete networks. One example of a simple optoplasmonic circuit, which combines the functions of light localization, frequency conversion and wavelength multiplexing, is schematically shown in Fig. 2a. It consists of two OMs (M1 and M2) with non-overlapping WG mode peaks (which can be tuned by OM morphology and material, see SI Fig. S1) and two plasmonic nanoantennas (D1 and D2) with a nanoantenna-coupled dipole sandwiched between the OMs. M1 is illuminated by a plane wave propagating along the z-axis. The incident light transverses M1 and is focused into the gap of the adjacent nanoantenna D1 where it generates a strong resonant and localized intensity enhancement at wavelength $\lambda_1$ (Fig. 2b,c). Nanoantenna D2 remains dark due to the efficient shielding through M2, which is in the off-resonance state at $\lambda_1$. The hot spot at D1 can also be formed by re-focusing the field radiated by another dipole source with emission wavelength $\lambda_1$ (similar to the case shown in Fig. 1a).

Acceptor molecules or QDs located in the gap of antenna D1 can be excited through the strongly enhanced localized field, provided that their absorption bands overlap with the hot spot wavelength $\lambda_1$. The excited acceptors eventually relax through emission of a photon. Due to the lack of coherency between the excitation and emission, the two processes can be treated independently (see Methods) (7). The enhancement of the radiative rate for a dipole located in the gap of D1 is plotted as function of the wavelength in Fig. 2d. The radiative rate is dramatically enhanced at defined wavelengths ($\lambda_{21}$ and $\lambda_{22}$) owing to the strong LDOS modification at D1 caused by the presence of M2 (M1 is off-resonance in this frequency range). The photons emitted in the gap of D1 can excite multiple WG modes in M2; and their resonant re-focusing in the nanoantenna D2 results in the formation of a multi-color nanoscale hot spot in D2 (Figs. 2e,f). The outlined optoplasmonic circuit could be implemented using a broadband emitter with a spectrum overlapping several WG modes (38, 39) or a cluster of narrow-band emitters such as size-selected QDs that allow dense packing without compromising their optical properties (40).

The underlying physical mechanisms behind the optoplasmonic wavelength multiplexing detailed in Fig. 2, where a localized single-color hot spot can create a multi-color hot-spot in a distant nanoantenna, can be naturally extended to design optoplasmonic demultiplexers. A schematic of a simple circuit element that performs wavelength-selective demultiplexing of emitted photons is shown in Fig. 3a. In this case, the strong resonant LDOS modification at the dipole position located between the OMs generates a double-peak radiative rate enhancement spectrum within a chosen frequency band (Fig. 3b). If the OMs are selected to provide spectrally-offset WG mode resonances, photons of different colors ($\lambda_1$ and $\lambda_2$) can be refocused into two spatially separated hot spots located at D1 ($\lambda_1$) and D2 ($\lambda_2$), respectively (Fig. 3c-e).

Integration of high-Q photonic elements into optoplasmonic circuits not only enables wavelength selectivity and efficient transfer of electromagnetic energy over longer distances but also offers the opportunity of cascaded signal amplification via efficient interaction of trapped photons with the gain medium inside the microcavity. This is at stark difference with conventional lossy nanoplasmonic components based on nanowires (13, 41) and channel plasmon polaritons (42) or SP loss compensation in planar device configurations (43, 44). The proposed optoplasmonic components also provide opportunities for realizing active nanoplasmonic circuit elements for field modulation and frequency switching, as the photon re-cycling in microcavities greatly enhances the sensitivity of light to small changes in refractive index (45). These properties are still missing in conventional nanoplasmonic circuitry due to the inherent weakness of the available material (e.g. thermo- electro- or magneto-optical) effects and the small propagating distance of SPs in metals (21).

While this work introduces the theoretical concepts of optoplasmonic superlenses, recent advances in nanofabrication technologies put the fabrication of these structures within reach. Our simulations predict significant radiative rate and light intensity enhancement with nanoantenna gaps sizes and position tolerances achievable by standard lithography (SI Fig. S4). Promising approaches to fabricate optoplasmonic networks include two-step electron-beam (46) and soft (47) lithography, template-assisted self-assembly (48), nanoassembly (16) and optical tweezers (49, 50) (see SI Fig. S5 for examples of possible realizations of optoplasmonic elements). It should be noted that OM shape and surface imperfections could have a detrimental impact on the spectral and energy transfer characteristics of optoplasmonic superlenses due to WG-modes splitting and multimode coupling. Both of these effects, which become more pronounced with the increase of the OM size (51), can, however, be alleviated by a proper engineering of the OM shape. Carefully designed OM shapes can be used to suppress some of the WG-modes and thus to rarefy the OM spectrum (e.g. higher-radial-order WG modes are suppressed in microring resonators (35, 36)). Successful realization of the proposed elements offers new opportunities for giant, highly frequency-sensitive and dynamically-controlled enhancement, transfer and routing of light on the nanoscale and could form a basis for new platforms for single-molecule imaging, bio(chemical) sensing and quantum information processing that interface photonic, plasmonic and electrical functionalities. Since electromagnetic signals in the proposed optoplasmonic networks may not only be detected but also launched and switched electrically, they could enable "dark" on-chip integrated circuits with all the coupling occurring in the near-field.

**Methods**

We calculate the radiative decay rate $\gamma_r$ of the dipole $\mathbf{p}$ at the emission wavelength $\lambda_{em}$ as the power fraction radiated into the far field by integrating the energy flux through the closed surface surrounding both the dipole and the optoplasmonic superlens. The non-radiative rate $\gamma_{nr}$ is found by integrating the energy flux through the closed surfaces enclosing individual lossy metal particles. The total decay rate $\gamma = \gamma_r + \gamma_{nr}$ can also be calculated from the total work per unit time that the electric field radiated by the dipole does on the dipole current: $\gamma \propto \mathrm{Im}\{\mathbf{E}(r_0, \lambda_{em}) \cdot \mathbf{p}\}$. The external quantum efficiency of optoplasmonic structures is defined as the ratio of the radiative and total decay rates: $q = \gamma_r / \gamma$. In turn, the excitation rate of an emitter with a transition dipole $\mathbf{p}$ is governed by the local field $\mathbf{E}_{exc}$ at the excitation wavelength $\lambda_{exc}$:

$\gamma_{exc} \propto |\mathbf{E}_{exc}(r_0, \lambda_{exc}) \cdot \mathbf{p}|^2$ (5, 6).

Generalized multi-particle Mie theory is used for all the calculations, which provides an exact analytical solution of Maxwell's equations for an arbitrary cluster of $L$ spheres(15). The total electromagnetic field scattered by the cluster is constructed as a superposition of partial fields scattered by each sphere. The incident, partial scattered and internal fields are expanded in the orthogonal basis of vector spherical harmonics represented in local coordinate systems associated with individual particles: $\mathbf{E}_{sc}^l = \sum_{(n)} \sum_{(m)} (a_{mn}^l \mathbf{N}_{mn} + b_{mn}^l \mathbf{M}_{mn}), \ l = 1,...L$. A matrix equation for the Lorenz-Mie multipole scattering coefficients ($a_{mn}^l, b_{mn}^l$) is obtained by imposing the continuity conditions for the tangential components of the electric and magnetic fields on the particles surfaces, using the translation theorem for vector spherical harmonics, and truncating the infinite series expansions to a maximum multipolar order $N$. Experimentally obtained Au refractive index values from Johnson and Christy (52) are used in the simulations.

**Acknowledgments.** The work was partially supported by the National Institutes of Health through grant 5R01CA138509-02 (BMR), the National Science Foundation through grants CBET-0853798 and CBET-0953121 (BMR) and the Army Research Laboratory Cooperative Agreement W911NF-06-2-0040 (BMR). Support from the EU COST Action MP0702 "Towards functional sub-wavelength photonic structures" (SVB) is also gratefully acknowledged.

**References**
1. Folan LM, Arnold S, & Druger SD (1985) Enhanced energy transfer within a microparticle. *Chem. Phys. Lett.* **118,** 322-327.
2. Vahala KJ (2003) Optical microcavities. *Nature* **424,** 839-846.
3. Englund D, Fattal D, Waks E, Solomon G, Zhang B, Nakaoka T, Arakawa Y, Yamamoto Y, & Vuckovic J (2005) Controlling the spontaneous emission rate of single quantum dots in a two-dimensional photonic crystal. *Phys. Rev. Lett.* **95,** 013904.

4. Badolato A, Hennessy K, Atature M, Dreiser J, Hu E, Petroff PM, & Imamoglu A (2005) Deterministic coupling of single quantum dots to single nanocavity modes. *Science* **308,** 1158-1161.
5. Aoki T, Dayan B, Wilcut E, Bowen WP, Parkins AS, Kippenberg TJ, Vahala KJ, & Kimble HJ (2006) Observation of strong coupling between one atom and a monolithic microresonator. *Nature* **443,** 671-674.
6. Halas NJ (2009) Connecting the dots: Reinventing optics for nanoscale dimensions. *Proc. Natl. Acad. Sci. USA* **106,** 3643-3644.
7. Bharadwaj P, Deutsch B, & Novotny L (2009) Optical antennas. *Adv. Opt. Photon.* **1,** 438-483.
8. Kühn S, Håkanson U, Rogobete L, & Sandoghdar V (2006) Enhancement of single-molecule fluorescence using a gold nanoparticle as an optical nanoantenna. *Phys. Rev. Lett.* **97,** 017402.
9. Kinkhabwala A, Yu Z, Fan S, Avlasevich Y, Mullen K, & Moerner WE (2009) Large single-molecule fluorescence enhancements produced by a bowtie nanoantenna. *Nature Photon.* **3,** 654-657.
10. Alu A & Engheta N (2008) Tuning the scattering response of optical nanoantennas with nanocircuit loads. *Nature Photon.* **2,** 307-310.
11. Muhlschlegel P, Eisler HJ, Martin OJF, Hecht B, & Pohl DW (2005) Resonant optical antennas. *Science* **308,** 1607-1609.
12. Cubukcu E, Kort EA, Crozier KB, & Capasso F (2006) Plasmonic laser antenna. *Appl. Phys. Lett.* **89,** 093120-093123.
13. Akimov AV, Mukherjee A, Yu CL, Chang DE, Zibrov AS, Hemmer PR, Park H, & Lukin MD (2007) Generation of single optical plasmons in metallic nanowires coupled to quantum dots. *Nature* **450,** 402-406.
14. Bergman DJ & Stockman MI (2003) Surface plasmon amplification by stimulated emission of radiation: quantum generation of coherent surface plasmons in nanosystems. *Phys. Rev. Lett.* **90,** 027402.
15. Chang DE, Sorensen AS, Hemmer PR, & Lukin MD (2006) Quantum optics with surface plasmons. *Phys. Rev. Lett.* **97,** 053002.
16. Barth M, Schietinger S, Fischer S, Becker J, Nusse N, Aichele T, Lochel B, Sonnichsen C, & Benson O (2010) Nanoassembled plasmonic-photonic hybrid cavity for tailored light-matter coupling. *Nano Lett.* **10,** 891-895.
17. Devilez A, Stout B, & Bonod N (2010) Compact metallo-dielectric optical antenna for ultra directional and enhanced radiative emission. *ACS Nano* **4,** 3390-3396.
18. Vernooy DW, Furusawa A, Georgiades NP, Ilchenko VS, & Kimble HJ (1998) Cavity QED with high-Q whispering gallery modes. *Phys. Rev. A* **57,** R2293.
19. Vollmer F, Arnold S, & Keng D (2008) Single virus detection from the reactive shift of a whispering-gallery mode. *Proc Natl Acad Sci USA* **105,** 20701-20704.
20. Barclay PE, Santori C, Fu K-M, Beausoleil RG, & Painter O (2009) Coherent interference effects in a nano-assembled diamond NV center cavity-QED system. *Opt. Express* **17,** 8081-8097.
21. Gramotnev DK & Bozhevolnyi SI (2010) Plasmonics beyond the diffraction limit. *Nature Photon.* **4,** 83-91.
22. Xu Y-l (1995) Electromagnetic scattering by an aggregate of spheres. *Appl. Opt.* **34,** 4573-4588.
23. Gopinath A, Boriskina SV, Feng N-N, Reinhard BM, & Negro LD (2008) Photonic-plasmonic scattering resonances in deterministic aperiodic structures. *Nano Lett.* **8,** 2423-2431.
24. Pishko SV, Sewell PD, Benson TM, & Boriskina SV (2007) Efficient analysis and design of low-loss whispering-gallery-mode coupled resonator optical waveguide bends. *J. Lightwave Technol.* **25,** 2487-2494.
25. Wylie JM & Sipe JE (1984) Quantum electrodynamics near an interface. *Phys. Rev. A* **30,** 1185.
26. Moskovits M, Tay LL, Yang J, & Haslett T (2002) in *Optical Properties of Nanostructured Random Media* (Springer-Verlag Berlin, Berlin), pp. 215-226.
27. Ward DR, Grady NK, Levin CS, Halas NJ, Wu Y, Nordlander P, & Natelson D (2007) Electromigrated nanoscale gaps for surface-enhanced Raman spectroscopy. *Nano Lett.* **7,** 1396-1400.
28. Kneipp K, Wang Y, Kneipp H, Perelman LT, Itzkan I, Dasari RR, & Feld MS (1997) Single molecule detection using surface-enhanced Raman scattering (SERS). *Phys. Rev. Lett.* **78,** 1667.
29. Alexander KD, Skinner K, Zhang S, Wei H, & Lopez R (2010) Tunable SERS in gold nanorod dimers

through strain control on an elastomeric substrate. *Nano Lett.* **10,** 4488–4493.
30. Falk AL, Koppens FHL, Yu CL, Kang K, de Leon Snapp N, Akimov AV, Jo M-H, Lukin MD, & Park H (2009) Near-field electrical detection of optical plasmons and single-plasmon sources. *Nature Phys.* **5,** 475-479.
31. Neutens P, Van Dorpe P, De Vlaminck I, Lagae L, & Borghs G (2009) Electrical detection of confined gap plasmons in metal-insulator-metal waveguides. *Nature Photon.* **3,** 283-286.
32. Ditlbacher H, Aussenegg FR, Krenn JR, Lamprecht B, Jakopic G, & Leising G (2006) Organic diodes as monolithically integrated surface plasmon polariton detectors. *Appl. Phys. Lett.* **89,** 161101-161103.
33. Heeres RW, Dorenbos SN, Koene B, Solomon GS, Kouwenhoven LP, & Zwiller V (2009) On-chip single plasmon detection. *Nano Lett.* **10,** 661-664.
34. Spillane SM, Kippenberg TJ, Painter OJ, & Vahala KJ (2003) Ideality in a fiber-taper-coupled microresonator system for application to cavity quantum electrodynamics. *Phys. Rev. Lett.* **91,** 043902.
35. Boriskina SV & Nosich AI (1999) Radiation and absorption losses of the whispering-gallery-mode dielectric resonators excited by a dielectric waveguide. *IEEE Trans. Microwave Theory Tech.* **47,** 224-231.
36. Hagness SC, Rafizadeh D, Ho ST, & Taflove A (1997) FDTD microcavity simulations: design and experimental realization of waveguide-coupled single-mode ring and whispering-gallery-mode disk resonators. *J. Lightwave Technol.* **15,** 2154-2165.
37. Yariv A, Xu Y, Lee RK, & Scherer A (1999) Coupled-resonator optical waveguide: a proposal and analysis. *Opt. Lett.* **24,** 711-713.
38. Biteen JS, Lewis NS, Atwater HA, Mertens H, & Polman A (2006) Spectral tuning of plasmon-enhanced silicon quantum dot luminescence. *Appl. Phys. Lett.* **88,** 131109-131103.
39. Ringler M, Schwemer A, Wunderlich M, Nichtl A, Kürzinger K, Klar TA, & Feldmann J (2008) Shaping emission spectra of fluorescent molecules with single plasmonic nanoresonators. *Phys. Rev. Lett.* **100,** 203002.
40. Klimov VI, Mikhailovsky AA, Xu S, Malko A, Hollingsworth JA, Leatherdale CA, Eisler HJ, & Bawendi MG (2000) Optical gain and stimulated emission in nanocrystal quantum dots. *Science* **290,** 314-317.
41. Fang Y, Li Z, Huang Y, Zhang S, Nordlander P, Halas NJ, & Xu H (2010) Branched silver nanowires as controllable plasmon routers. *Nano Lett.* **10,** 1950-1954.
42. Volkov VS, Bozhevolnyi SI, Devaux E, Laluet J-Y, & Ebbesen TW (2007) Wavelength selective nanophotonic components utilizing channel plasmon polaritons. *Nano Lett.* **7,** 880-884.
43. Ambati M, Nam SH, Ulin-Avila E, Genov DA, Bartal G, & Zhang X (2008) Observation of stimulated emission of surface plasmon polaritons. *Nano Lett.* **8,** 3998-4001.
44. Noginov MA, Podolskiy VA, Zhu G, Mayy M, Bahoura M, Adegoke JA, Ritzo BA, & Reynolds K (2008) Compensation of loss in propagating surface plasmon polariton by gain in adjacent dielectric medium. *Opt. Express* **16,** 1385-1392.
45. Almeida VR, Barrios CA, Panepucci RR, & Lipson M (2004) All-optical control of light on a silicon chip. *Nature* **431,** 1081-1084.
46. Curto AG, Volpe G, Taminiau TH, Kreuzer MP, Quidant R, & van Hulst NF (2010) Unidirectional emission of a quantum dot coupled to a nanoantenna. *Science* **329,** 930-933.
47. Smythe EJ, Dickey MD, Whitesides GM, & Capasso F (2008) A technique to transfer metallic nanoscale patterns to small and non-planar surfaces. *ACS Nano* **3,** 59-65.
48. Yan B, Thubagere A, Premasiri WR, Ziegler LD, Dal Negro L, & Reinhard BM (2009) Engineered SERS substrates with multiscale signal enhancement: nanoparticle cluster arrays. *ACS Nano* **3,** 1190-1202.
49. Chiou PY, Ohta AT, & Wu MC (2005) Massively parallel manipulation of single cells and microparticles using optical images. *Nature* **436,** 370-372.
50. Grier DG (2003) A revolution in optical manipulation. *Nature* **424,** 810-816.
51. Matsko AB & Ilchenko VS (2006) Optical resonators with whispering-gallery modes-part I: basics. *IEEE J. Select. Top. Quantum Electron.* **12,** 3-14.
52. Johnson PB & Christy RW (1972) Optical constants of the noble metals. *Phys. Rev. B* **6,** 4370.

53. Armani AM & Vahala KJ (2006) Heavy water detection using ultra-high-Q microcavities. *Opt. Lett.* **31,** 1896-1898.
54. Johnson BR (1993) Theory of morphology-dependent resonances: shape resonances and width formulas. *J. Opt. Soc. Am. A* **10,** 343-352.
55. Teraoka I & Arnold S (2009) Resonance shifts of counterpropagating whispering-gallery modes: degenerate perturbation theory and application to resonator sensors with axial symmetry. *J. Opt. Soc. Am. B* **26,** 1321-1329.
56. Chang DE, oslash, rensen AS, Hemmer PR, & Lukin MD (2006) Quantum Optics with Surface Plasmons. *Phys. Rev. Lett.* **97,** 053002.
57. Lipomi DJ, Kats MA, Kim P, Kang SH, Aizenberg J, Capasso F, & Whitesides GM Fabrication and replication of arrays of single- or multicomponent nanostructures by replica molding and mechanical sectioning. *ACS Nano* **4,** 4017-4026.

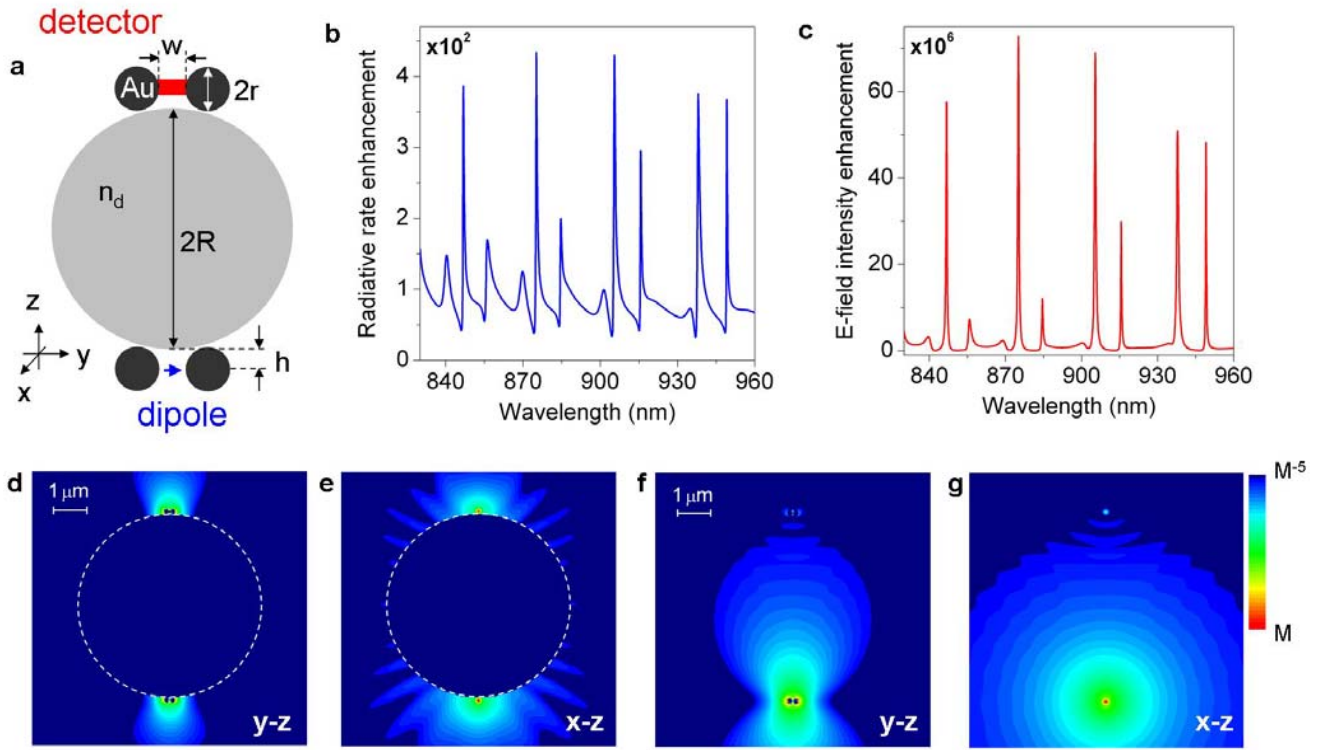

**Figure 1. Resonant amplifying superlens.** (a) A schematic of the optoplasmonic superlens composed of a polystyrene microsphere and two Au nanodimer antennas ($R$=2.8µm, $n_d$=1.59, $r$=75nm, $w$=25nm, $h$=80nm). The structure is excited by an electric dipole shown as the blue arrow. (b) Radiative rate enhancement of the dipole (over the free-space value, $\gamma_r / \gamma_r^0$) as a function of wavelength. (c) Electric field intensity enhancement in the gap of the acceptor antenna (over the value generated at the same position by a free-space dipole, $|E|^2 / |E_0|^2$). (d,e) Electric field intensity distribution in the y-z (d) and x-z (e) planes (log scale) at one of the resonant peaks in b,c (λ=905.4nm). (f,g) Electric field intensity distribution at the same wavelength in the y-z (f) and x-z (g) planes (log scale) in the absence of the microsphere.

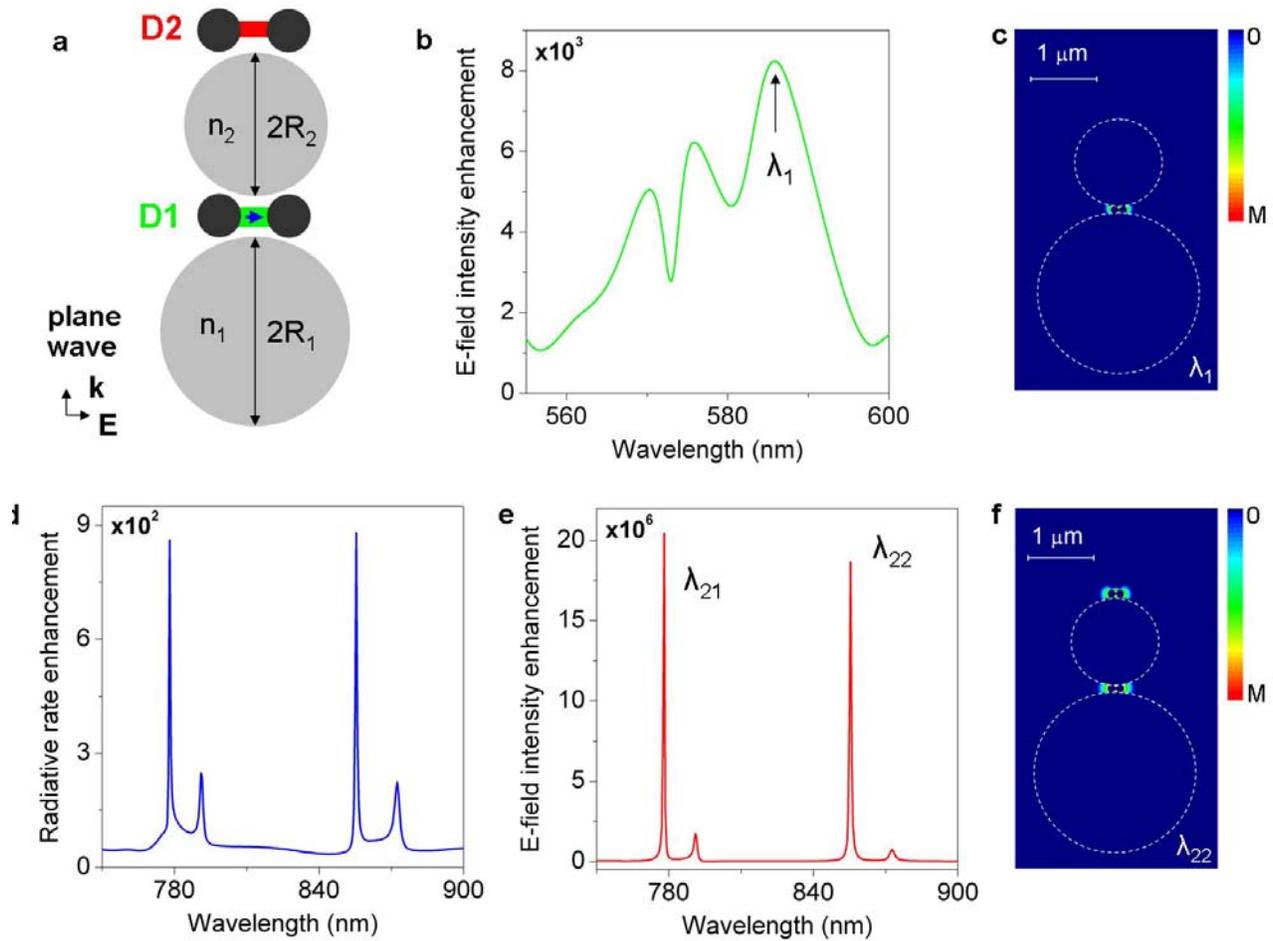

**Figure 2. Light in-coupling, frequency conversion and multiplexing in optoplasmonic circuits.** (a) A schematic of the optoplasmonic multiplexer ($R_1$=1.2µm, $n_1$=1.45, $R_2$=0.65µm, $n_2$=2.4, $r$=55nm, $w$=20nm, $h$=60nm). (b) Enhancement (over the free-space value) of the electric field intensity detected at D1 under the plane wave excitation. (c) Electric field intensity distribution (linear scale) in the multiplexer at $\lambda_1$=585nm under the plane wave illumination. (d) Radiative rate enhancement of the dipole at D1 as a function of wavelength. (e) Electric field intensity enhancement detected at D2 under the excitation by the emitting dipole at D1 as a function of wavelength. (f) Electric field intensity distribution (linear scale) in the multiplexer at $\lambda_{22}$=855.5nm under the illumination by the emitting dipole at D1.

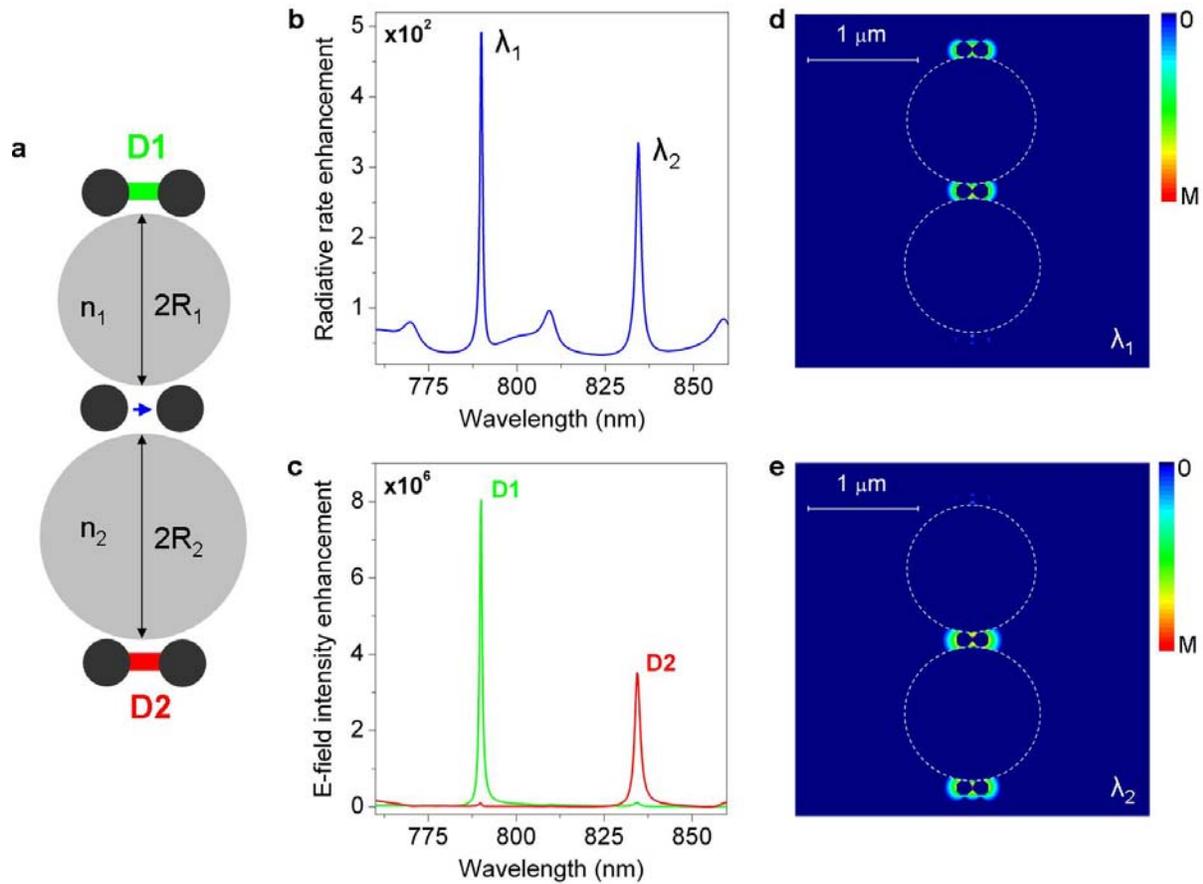

**Figure 3. Optoplasmonic frequency demultiplexer**. (a) A schematic of the optoplasmonic demultiplexer: a field generated by the emitting dipole (shown as a blue arrow) is monitored in the gaps of two nanoantennas ($R_1$=0.57µm, $n_1$=2.4, $R_2$=0.6µm, $n_2$=2.4, $r$=65nm, $w$=25nm, $h$=70nm). (b) Dipole radiative rate enhancement as a function of wavelength. (c) The frequency spectra of the electric field intensity enhancement detected at D1 (green) and D2 (red). (d,e) Electric field intensity distributions (linear scale)
in the demultiplexer at $\lambda_1$=790nm (d) and $\lambda_2$=834nm (e).

## Supplementary information

High-Q optical microcavities can strongly modify the LDOS and thus can be used for the frequency- and position-selective enhancement(suppression) of the radiative rates of emitters either evanescently coupled to or embedded inside the cavities. The enhancement is characterized by the Purcell factor, and can be large in cavities that support optical modes with high quality factors and small mode volumes(3, 4, 18, 20, 53). Fig. S1 illustrates this effect for a model structure composed of a single dipole emitter with a transition moment oriented along the y-axis, which is evanescently coupled to a microsphere via a 1nm-wide gap (Fig. S1**a**). The spectra of the microcavity-mediated dipole radiative rate enhancement shown in Fig. S1**b** feature a series of sharp peaks that correspond to the dipole emission coupling to the whispering-gallery (WG) modes inside the microspheres. The spectral positions of the resonances are determined by the morphology of the microcavity (its size and material composition), and the resonance linewidths are inversely proportional to the WG mode Q-factors.

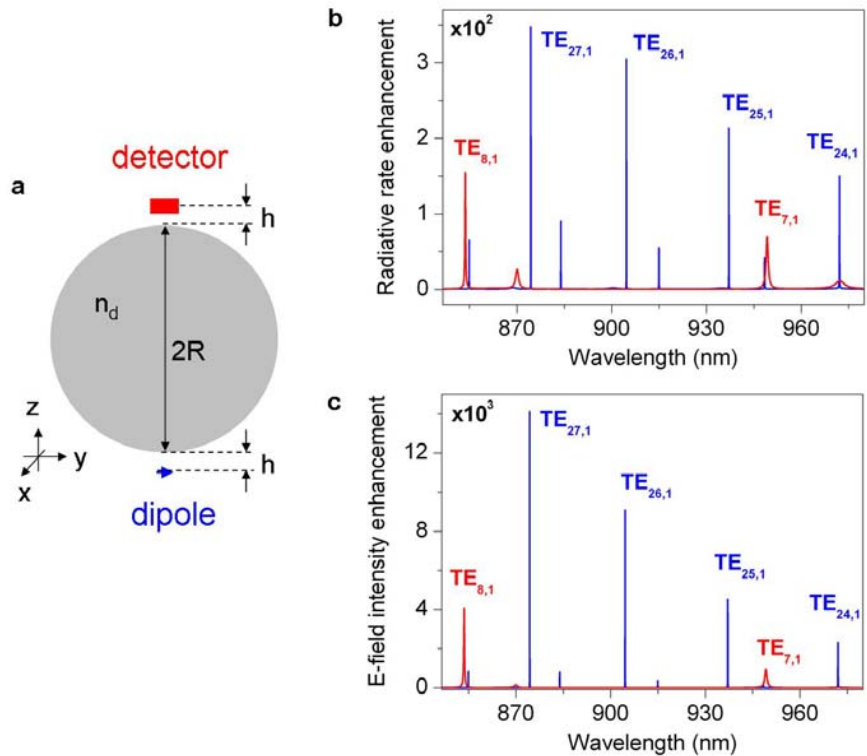

**Supplementary Figure S1**. **a**, A schematic of the microsphere excited by a dipolar source (blue arrow, $h$=1nm) and the position of the detector monitoring the electric field (red bar). **b**, The dipole radiative rate enhancement (over the free-space value) as a function of wavelength and the sphere morphology (blue: $R$=2.8μm, $n_d$=1.59, $h$=1nm; red: $R$=0.65μm, $n_d$=2.4, $h$=1nm). The resonant peaks correspond to the excitation of the WG modes in the microspheres. **c**, Electric field intensity enhancement at the detector position (over the value generated by a free-space dipole at the same position).

In general, each WG mode in a microsphere can be specified by four indices: $n$, the radial order (the number of peaks in the intensity profile along the radial direction); $m$, the azimuthal mode index (the number of field variations along the sphere equator); $l$, the number of waves in a cyclic orbit ($l - |m| + 1$ is equal to the number of peaks in the intensity profile of the mode along the meridian); and polarization, TE (transverse-electric) or TM (transverse magnetic)(54, 55). Each WG mode is multiple-degenerate in frequency with $m$ taking the values $l, l-1 ... -l$ for each $n$. Our simulations show that the most pronounced spectral peaks in Fig. S1**b** correspond to the excitation of TE-polarized WG modes with $n = 1$ and $m = l$ in the microsphere (see Fig. S2), which are indexed as TE$_{m,1}$ modes in Fig. S1**b**. The spectra of the electric field intensity monitored on the opposite side of the microspheres demonstrate narrow-peak resonant enhancement produced by the evanescent tails of the WG modes (Fig. S1**c**). However, as the optical leakage of the WG modes occurs along the whole circumference of the sphere, light re-focusing into a nanoscale volume cannot be achieved. Typical electric field intensity distributions of TE$_{nm}$ modes in microsphere resonators are shown in Fig. S2.

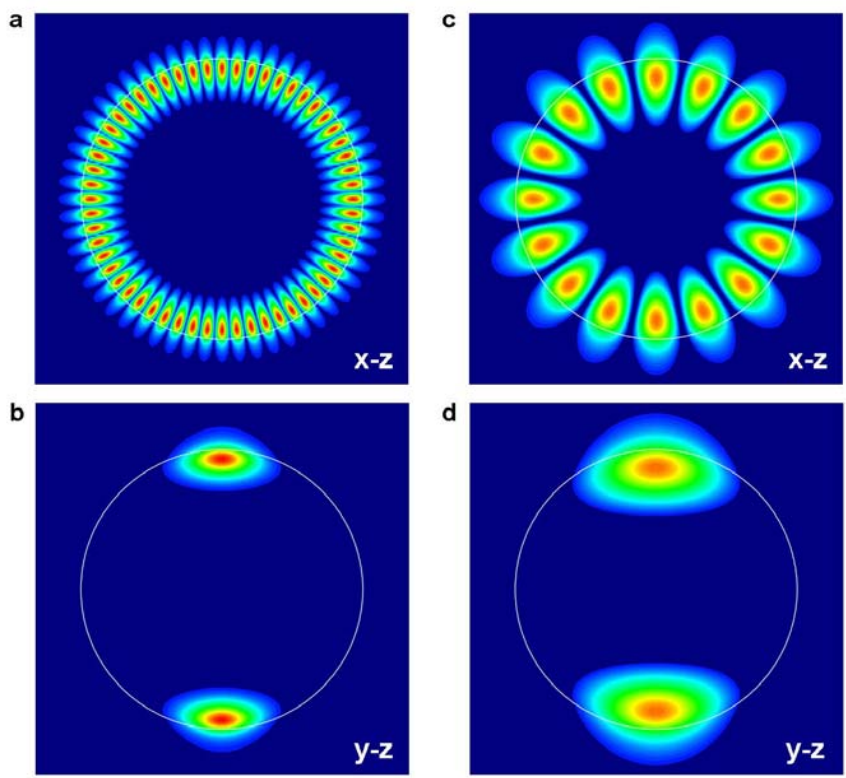

**Supplementary Figure S2**. **a-d**, Field intensity distributions in the x-z (**a,c**) and y-z (**b,d**) planes of the TE$_{28,1}$ whispering gallery mode in a larger polystyrene microsphere (**a,b**, $R$=2.8μm, $n_d$=1.59, $\lambda_{res}$=846.12nm, Q=1.05·10$^5$) and of the TE$_{8,1}$ WG mode in a smaller TiO$_2$ microsphere (**c,d**, $R$=650nm, $n_d$=2.4, $\lambda_{res}$=853.62nm, Q=3204).

Strong coherent coupling of the fields of nanoscale emitters to surface plasmon resonances in noble-metal nanostructures also results in strong modification of their radiative and non-radiative decay rates(7, 8, 56). In this case, the Purcell enhancement of dipole emission is enabled by the strongly reduced effective mode volume for the photons, while the Q-factors of plasmonic nanostructures are limited by high dissipative losses in metals at optical wavelengths. A schematic of a nanoantenna-coupled dipole source is shown in Fig. S3**a** together with the second acceptor nanoantenna, which is separated from the donor one by a micron-scale distance L. The spectra of the nanoantenna-mediated emitter radiative rate enhancement feature a single broad peak that corresponds to the excitation of the bonding dipole plasmon resonance of the nanodimer (with the peak wavelength determined by the Au nanoparticles radii and the antenna gap width, see Fig. S3**b**). The spectra of the field intensity measured in the gap of the acceptor nanodimer also feature a single broad peak (Fig. S3**c**). However, a free-space energy transfer between two plasmonic nanoantennas is not efficient as most of the energy radiated by the donor nanoantenna cannot be re-captured by the acceptor antenna (see Fig. 1**f,g**).

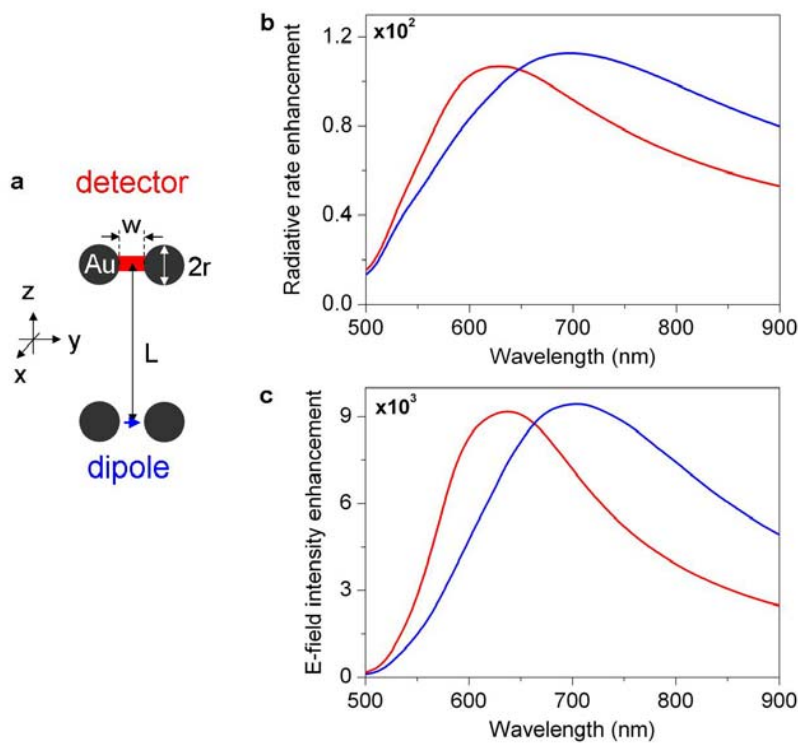

**Supplementary Figure S3**. **a**, A schematic of two Au nanodimer antennas showing a position of the dipole emitter and the field intensity monitor. **b**, Radiative rate enhancement (over the free-space value) of the dipole positioned in the gap of the donor antenna as a function of wavelength and antenna geometry (red: $r$=65nm, $w$=25nm, $L$=652nm; blue: $r$=75nm, $w$=25nm, $L$=2.802µm). **c**, Electric field intensity enhancement in the gap of the acceptor antenna (over the value generated by a free-space dipole at the same position).

The results of our simulations shown in Fig.1 of the main text demonstrate that a combination of the high Q-factors of the microcavity modes and the strong field localization in the gap of plasmonic antennas result in the resonant increase of the dipole radiative rates over the values achievable either by using a microcavity or an antenna alone. In the optoplasmonic superlens configuration considered in this Letter, plasmonic antennas interact with the microcavity modes via the exponentially decaying tails of their evanescent fields (Fig. S4**a**). Therefore, the interaction is distance-dependent and, as shown in Fig.S4**b**, the amplitude of the effect reduces with the antenna-cavity separation. However, the data in Fig. S4**b** demonstrate a high tolerance of the values of the radiative rate enhancement to the variations (up to tens of nanometers) in the width of the antenna-cavity gap. Furthermore, our calculations show that coupling of the antenna to the microcavity not only can enhance the dipole radiative rate but also increases its external quantum efficiency (Fig. S4**c**). Finally, the efficiency of the light re-focusing in the acceptor antenna is robust to small variations in the antenna-microcavity coupling distance (Fig. S4**d,e**).

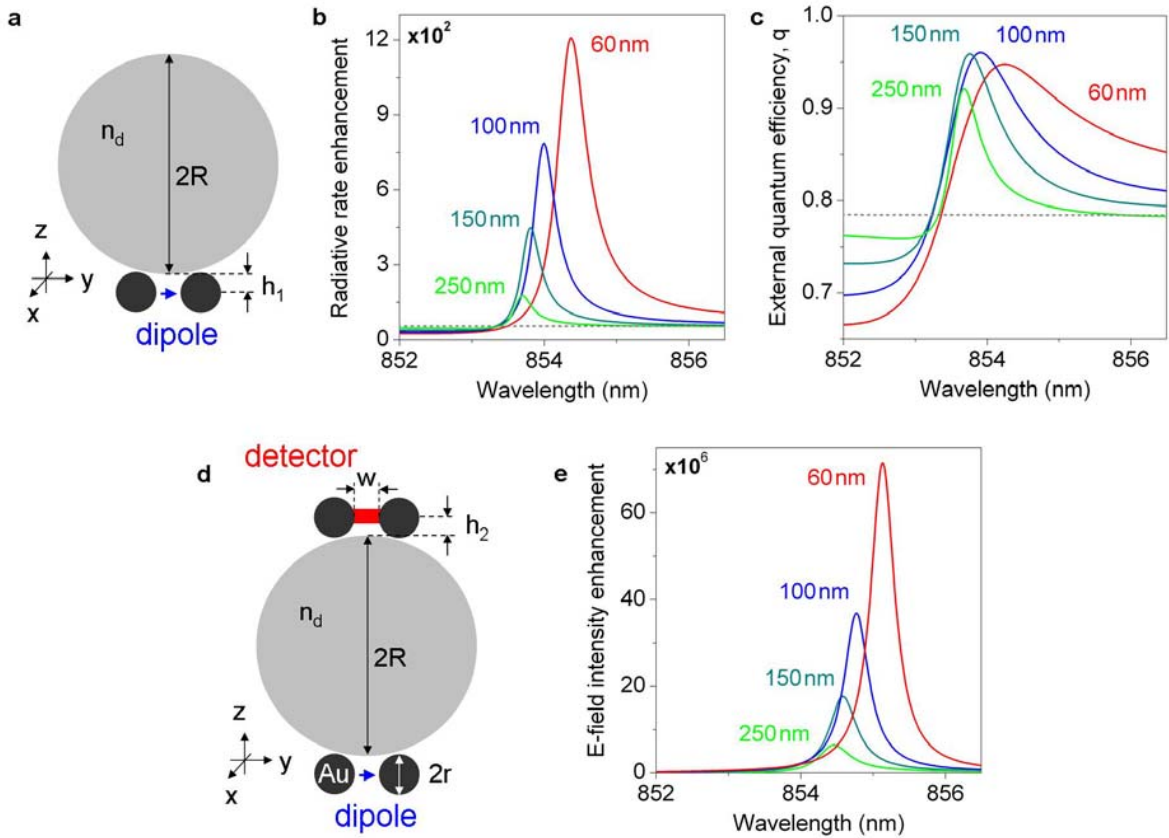

**Supplementary Figure S4**. **a**, A schematic of an Au nanodimer antenna coupled to a microcavity via an airgap of width $h_1$ ($R$=650nm, $n_d$=2.4, $r$=55nm, $w$=20nm) **b**, Radiative rate enhancement of a dipole positioned in the center of the antenna gap as a function of the wavelength around one of the WG-mode peaks and the antenna-microcavity separation $h_1$. The labels indicate the values of $h_1$, and the dashed line shows the antenna-mediated dipole radiative rate enhancement in the absence of the microcavity. **c**, External quantum efficiency $q$ of the antenna-microcavity structure. The dashed line shows the corresponding value of the nanodimer antenna. **d**, A schematic of the optoplasmonic superlens with the acceptor antenna separated from the microcavity via the airgap of width $h_2$. e, Electric field intensity enhancement in the gap of the acceptor antenna as a function of wavelength and $h_2$ ($h_1$=60nm, the labels indicate the values of $h_2$).

As mention in the main text, a number of nano-fabrication techniques can be used to fabricate the proposed bright- and dark-field optoplasmonic superlenses and on-chip integrated optoplasmonic networks. Some of the examples of possible realizations of the optoplasmonic components and circuits are shown in Fig. S5 and can include not only spherical particles or microcavities but also planar microdisk, microring and microtoroid resonators as well as bow-tie and other types of dimer gap antennas and metal nanoclusters.

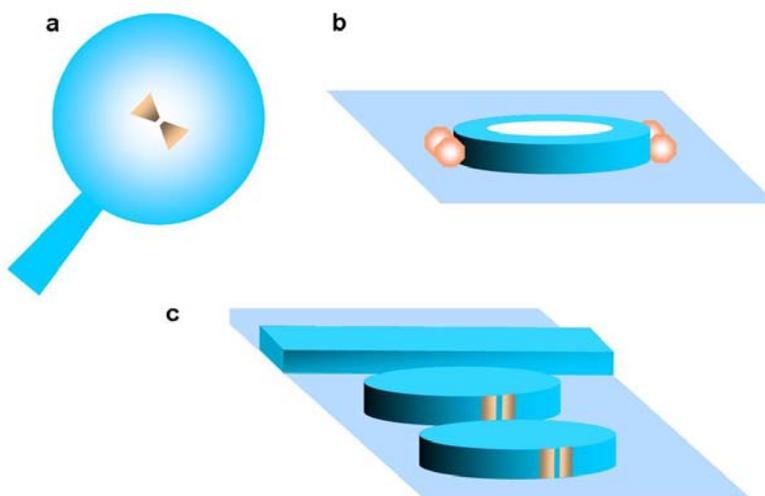

**Supplementary Figure S5. a**, Plasmonic nanoantenna(s) can be transferred to a surface of a dielectric microsphere(47). **b**, On-chip microdisk or microring resonators decorated with nanosphere dimers can be fabricated by a combination of lithography and nanomanipulation(16). **c**, Planar networks of dielectric (semiconductor) microdisks laterally coupled to plasmonic nanoantennas and photonic nanowire waveguides are also amenable to fabrication by two-step conventional and soft lithographic techniques(46, 57).

## References

1   Barclay, P. E. *et al.* Coherent interference effects in a nano-assembled diamond NV center cavity-QED system. *Opt. Express* **17**, 8081-8097 (2009).

2   Englund, D. *et al.* Controlling the spontaneous emission rate of single quantum dots in a two-dimensional photonic crystal. *Phys. Rev. Lett.* **95**, 013904 (2005).

3   Badolato, A. *et al.* Deterministic coupling of single quantum dots to single nanocavity modes. *Science* **308**, 1158-1161 (2005).

Vernooy, D. W. *et al.* Cavity QED with high-Q whispering gallery modes. *Phys. Rev. A* **57**, R2293 (1998).

5    Armani, A. M. & Vahala, K. J. Heavy water detection using ultra-high-Q microcavities. *Opt. Lett.* **31**, 1896-1898 (2006).

6    Johnson, B. R. Theory of morphology-dependent resonances: shape resonances and width formulas. *J. Opt. Soc. Am. A* **10**, 343-352 (1993).

7    Teraoka, I. & Arnold, S. Resonance shifts of counterpropagating whispering-gallery modes: degenerate perturbation theory and application to resonator sensors with axial symmetry. *J. Opt. Soc. Am. B* **26**, 1321-1329 (2009).

8    Kühn, S., Håkanson, U., Rogobete, L. & Sandoghdar, V. Enhancement of single-molecule fluorescence using a gold nanoparticle as an optical nanoantenna. *Phys. Rev. Lett.* **97**, 017402 (2006).

9    Bharadwaj, P., Deutsch, B. & Novotny, L. Optical antennas. *Adv. Opt. Photon.* **1**, 438-483 (2009).

10    Chang, D. E. *et al.* Quantum Optics with Surface Plasmons. *Phys. Rev. Lett.* **97**, 053002 (2006).

11    Smythe, E. J., Dickey, M. D., Whitesides, G. M. & Capasso, F. A technique to transfer metallic nanoscale patterns to small and non-planar surfaces. *ACS Nano* **3**, 59-65 (2008).

12    Barth, M. *et al.* Nanoassembled plasmonic-photonic hybrid cavity for tailored light-matter coupling. *Nano Lett.* **10**, 891-895 (2010).

13    Curto, A. G. *et al.* Unidirectional emission of a quantum dot coupled to a nanoantenna. *Science* **329**, 930-933 (2010).

14    Lipomi, D. J. *et al.* Fabrication and replication of arrays of single- or multicomponent nanostructures by replica molding and mechanical sectioning. *ACS Nano* **4**, 4017-4026.
16